\documentclass[a4paper,10pt]{article}

\usepackage{amsmath}
\usepackage{amsthm}
\usepackage{graphicx}

\usepackage[algoruled,english,linesnumbered,lined]{algorithm2e} %

\newtheorem{theorem}{Theorem}

\newcounter{todo}

\newcommand{\rank}{p}

\newcommand{\seqGilt}[2]{\left\langle #1\gilt #2\right\rangle}
\newcommand{\Id}[1]{\ensuremath{\mathit{#1}}}
\newcommand{\ceil}[1]{\left\lceil #1\right\rceil}
\newcommand{\floor}[1]{\left\lfloor #1\right\rfloor}

\newcommand{\gilt}{:}

\newcommand{\setGilt}[2]{\left\{ #1\gilt #2\right\}}

\newcommand{\realrange}[2]{\left[#1, #2\right]}

\newcommand{\unitrange}[2]{\realrange{0}{1}}

\newcommand{\Th}[1]{\Theta\!\left( #1\right)}

\newcommand{\llabel}[1]{\label{\labelprefix:#1}}
\newcommand{\labelprefix}{} %

\newcommand{\discussionsize}{\small}

\newenvironment{code}{\noindent%
\begin{tabbing}%
\hspace{2em}\=\hspace{2em}\=\hspace{2em}\=\hspace{2em}\=\hspace{2em}\=%
\hspace{2em}\=\hspace{2em}\=\hspace{2em}\=\hspace{2em}\=\hspace{2em}\=%
\kill}{\end{tabbing}}

\newcommand{\labelcommand}{}
\newcommand{\captiontext}{}
\newsavebox{\codeparam}
\newcounter{lineNumber}

\newdimen\endofsize\endofsize=0.5em
\def\endofbeweis{~\quad\hglue\hsize minus\hsize
                 \hbox{\vrule height \endofsize width
\endofsize}\par}
{\par\vspace{1.0ex}\noindent{\bf Example}\ }%
{\par\vspace{0.0ex}\noindent}

\begin{document}

\title{Towards Optimal Range Medians}

\author{Beat Gfeller \\ {\small ETH Zurich, Switzerland} \\ \small{ \texttt{gfeller@inf.ethz.ch}} \and Peter Sanders\thanks{Partially supported by DFG grant SA 933/3-1.}  \\ {\small Universit\"at Karlsruhe, Germany} \\ {\small \texttt{sanders@ira.uka.de}} \smallskip}

\maketitle

\begin{abstract}
  We consider the following problem: given an unsorted array of $n$
  elements, and a sequence of intervals in the array,
  compute the median in each of
  the subarrays defined by the intervals.  We describe a simple
  algorithm which uses $O(n)$ space and needs $O(n\log k + k\log n)$ time
  to answer the first $k$  queries.
  This improves previous algorithms by a logarithmic factor and matches a lower bound
  for $k=O(n)$.
  Since the algorithm decomposes the range of element values rather than the array,
  it has natural generalizations to higher dimensional problems -- it reduces a range median query
  to a logarithmic number of range counting queries. 
\end{abstract}

\section{Introduction and Related Work}

The classical problem of finding the \emph{median} is to find the element of rank $\lceil n/2 \rceil$ in an unsorted array of $n$ elements.\footnote{An element has rank $i$ if it is the $i$-th element in some sorted order. Actually, any specified rank might be of interest. We restrict 
ourselves to the median to simplify notation but a generalization to arbitrary ranks will be straightforward for all our results.}
Clearly, the median can be found in $O(n\log n)$ time by sorting the elements. However, a classical algorithm 
finds the median in $O(n)$ time \cite{DBLP:conf/stoc/BlumFPRT72}, which is asymptotically optimal.

More recently, the following generalization, called the \emph{Range Median Problem (RMP)}, has been considered \cite{DBLP:journals/njc/KrizancMS05,DBLP:conf/esa/Har-PeledM08}:

\noindent{\bf Input:} An unsorted array $A$ with $n$ elements, each
having a \emph{value}. Furthermore, a sequence of $k$ \emph{queries}
$Q_1,\ldots,Q_k$, each defined as an interval $Q_i =[l_i,r_i]$.
In general, the sequence is given in an online
fashion, 
we want an answer after every query, and $k$ is not known in advance.

\noindent{\bf Output:} A sequence $x_1,\ldots,x_k$ of values, where
$x_i$ is the median of the elements in $A[l_i,r_i]$.  Here, $A[l,r]$
denotes the set of all elements whose index in $A$ is at least
$l$ and at most $r$.

This RMP naturally fits into a larger group of problems, in which an
unsorted array is given, and in a query one wants to compute a certain function of all the elements in a
given interval. 
Instead of the median, natural candidates for such a function are:
\begin{itemize}
 \item Sum: this problem can be trivially solved in $O(n)$ preprocessing time and $O(1)$ query time by computing prefix sums.
\item Semigroup operator: this problem is significantly more difficult than the sum. However, there exists a very efficient solution: for any constant $c$, a preprocessing in $O(nc)$ time and space allows to answer queries in $O(\alpha_c(n))$ time,
where $\alpha_k$ is a certain functional inverse of the Ackerman function \cite{802185}. A matching lower bound was given in \cite{DBLP:journals/siamcomp/Yao85a}.
\item Maximum, Minimum: This is a special case of a semigroup operator, for which the problem can be solved slightly more efficiently: $O(n)$ preprocessing time and space is sufficient to allow $O(1)$ time queries (see e.g. \cite{DBLP:journals/tcs/BenderF04}).
\end{itemize}
In addition to being a natural extension of the median problem, the RMP has some applications in practice, namely obtaining a ``typical'' element in a given time series out of a given time interval \cite{DBLP:conf/esa/Har-PeledM08}.

Natural special cases of the RMP are an \emph{offline} variant, where all queries are given 
in a batch and a variant where we want to do all \emph{preprocessing up front} and are then interested in
good worst case bounds for answering a single query.

The authors of \cite{DBLP:conf/esa/Har-PeledM08} give a solution of
the RMP which requires $O(n \log k + k\log n \log k)$ time and
$O(n\log k)$ space. In addition, they give a lower bound of
$\Omega(n\log k)$ time for comparison-based algorithms.
They basically use a one-dimensional range tree over the input array, where each inner node corresponds to a
subarray defined by an interval. Each such subarray is  sorted, and stored with the node. 
A range median query then corresponds to selecting the median from $O(\log k)$  sorted subarrays (whose union is the queried subarray) of total length $O(n)$, which requires $O(\log n \log k)$ time. 
The main difficulty of their approach is to show that  the subarrays need  not be fully sorted, but only presorted in a particular way, which 
reduces the construction time of the tree from $O(n\log n)$ to $O(n \log k)$.

Concerning the preprocessing variant of the RMP,
\cite{DBLP:journals/njc/KrizancMS05} give a data structure and answers queries in $O(\log n)$ time,
which uses $O(n \log^2 n /\log\log n)$ space
They do not analyze the required preprocessing time, but it is clearly at least as large as the required space in words.
Moreover, they give a structure which uses only $O(n)$ space, but query time $O(n^\epsilon)$ for arbitrary $\epsilon>0$.
 Another data structure given in \cite{DBLP:conf/sofsem/Petersen08a}
can answer queries in $O(1)$ time and uses $O(n^2 \log^{(p)} n / \log n)$ space, where $p$ is an arbitrary integer, and $\log^{(p)} n$ is the $p$ times iterated logarithm of $n$.\footnote{Note that the data structures in \cite{DBLP:journals/njc/KrizancMS05,DBLP:conf/sofsem/Petersen08a} 
work only for a specific quantile (e.g. the median), which must be the same for all queries.}

\subsubsection*{Our results.}
First, in Section~\ref{s:pointermachine} we give an algorithm for the
pointer-machine model which solves the RMP in $O(n\log k + k\log n)$
time and $O(n\log k)$ space. This improves the running time of $O(n
\log k + k\log n \log k)$ reported in
\cite{DBLP:conf/esa/Har-PeledM08} for $k \in \omega(n/\log n)$.  Our
algorithm is also considerably simpler.  The idea is to reduce a
range median query to a logarithmic number of related range counting
queries by recursively partitioning the values in array $A$ in a tree
in a fashion similar to Quicksort.  The final time bound is achieved
using the technique of Fractional Cascading.
In Section~\ref{ss:lowerbound}, we explain why our algorithm is optimal for $k\in O(n)$ and at most
$\Omega(\log n)$ from optimal for $k\in \omega(n)$.

Section~\ref{s:RAM} achieves linear space in the RAM model using
techniques from succinct data structures -- the range counting
problems are reduced to rank computations in bit arrays. To achieve
the desired bound, we compress the recursive subproblems in such a way
that the bit arrays remain dense at all times.

The latter algorithm can be easily modified to obtain a data structure using $O(n)$ space, can be constructed in $O(n \log n)$ time, and allows to answer arbitrary range median queries (or an arbitrary rank, which may be specified online together with the query interval) in $O(\log n)$ time. 
Note that the previously best linear-space data structure required $O(n^\epsilon)$ query time \cite{DBLP:journals/njc/KrizancMS05}.

After a few remarks on generalizations for higher dimensional inputs
in Section~\ref{s:highd}, we discuss dynamic variants of the RMP problem in Section~\ref{s:dynamic}, before Section~\ref{s:conclusions} concludes with a
summary and some open problems.

\section{A Pointer Machine Algorithm}
\label{s:pointermachine}

\begin{figure}
\begin{center}
\includegraphics{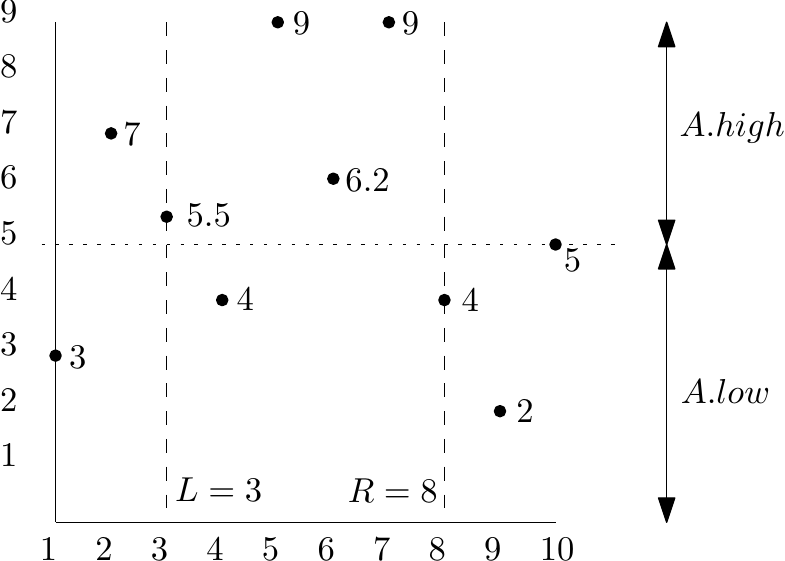}
\end{center}
\caption{An example for the array $A=[3,7,5.5,4,9,6.2,9,4,2,5]$.
The median value used to split the set of points is $5$. For the query $L=3,R=8$, there are two elements inside $A.low[L,R]$ and four elements in $A.high[L,R]$. Hence, the median of $A[L,R]$ is the element of rank $ \rank = \lceil(2+4)/2 \rceil - 2 = 1$ in $A.high[L,R]$. }
\label{fig:key-observation}
\end{figure}

Our algorithm is based is the following key observation
(see also Figure~\ref{fig:key-observation}): Suppose that we partition
the elements in array $A$ of length $n$ into two smaller arrays:
$A.low$ which contains all elements with the $n/2$
smallest\footnote{To simplify
  notation we ignore some trivial rounding issues and also sometimes
  assume that all elements have unique values. This is without loss of
  generality because we could artificially expand the size $A$ to the next power of two and
  because we can use the position of an element in $A$ to break ties in element comparisons.} values in $A$, and $A.high$ which contains all elements with the
$n/2$ largest values. The elements in $A.low$ and $A.high$ are sorted
by their index in $A$, and each element $e$ in $A.low$ and $A.high$ is
associated with its index $e.i$ in the original input array, and its value $e.v$.  Now, if
we want to find the element of rank $\rank$ in the subarray $A[L,R]$,
we can do the following: We count the number $m$ of elements in $A.low$
which are contained in $A[L,R]$.  To obtain $m$, we can do a binary
search for both $L$ and $R$ in $A.low$ (using the $e.i$ fields).  If
$\rank\leq m$, then the element of rank $\rank$ in $A[L,R]$ is
 the element of rank $\rank$ in $A.low[L,R]$. Otherwise, the
element of rank $\rank$ is  the element of rank $\rank-m$ in
$A.high[L,R]$.

Hence, using the partition of $A$ into $A.low$ and $A.high$, we can reduce the problem of finding an element of a given rank in array $A[L,R]$
to the same problem, but on a smaller array (either $A.low[L,R]$ or $A.high[L,R]$). Our algorithm applies this reduction recursively.

\paragraph*{Algorithm overview.}
The basic idea is therefore to subdivide the $n$ elements in the array
into two parts of (almost) equal size by computing the median of
their values and using it to split the list into a list of
the $n/2$
elements with smaller values and a list of the $n/2$
elements with larger values.  The two parts are recursively subdivided
further, but only when required by a query.
To answer a range median query, we determine in which of the two parts
the element of the desired rank lies (initially, this rank corresponds
to the median, but this may change during the search).  Once this is
known, the search continues recursively in the appropriate part until
a trivial problem of constant size  is encountered.

We will show that the total work involved in splitting the subarrays
is $O(n\log k)$ and that the search required for any query can be completed in $O(\log n)$ time using Fractional
Cascading \cite{fractionalcascading}.
Hence, the total running time is $O(n \log k + k\log n) $.

\paragraph*{Detailed description and analysis.}
\begin{algorithm}
\caption{Query($A,L,R,\rank$)} \label{alg:query}
\dontprintsemicolon
{\bf Input:} range select data structure $A$ of elements, query range $[L,R]$, desired rank $\rank$\;
\lIf{$|A|=1$}{\Return $A[1]$} \;
\uIf{$A.{low}$ is undefined}{
  Compute median $x$ value of the pairs in $A$ \;
  $A.{low}\hspace*{1.75mm} \  := \seqGilt{e\in A}{e.v\leq x}$\;
  $A.{high} \  := \seqGilt{e\in A}{e.v> x}$

  $\{$ $\seqGilt{e\in A}{ \mathcal{Q} }$ is an array containing all elements $e$ of $A$ satisfying the given condition $\mathcal{Q}$, ordered as in $A$ $\}$ 
}
\{ Find($A,q$) returns  $\max\setGilt{j}{A[j].i\leq q}$ (with Find$(A,0)=0$)  \} \; 
$ l := $ Find($A.{low},L-1)$  \tcp*{\# of low elements left of $L$}\;
$ r := $ Find($A.{low},R)$  \tcp*{\# of low elements up to $R$}\;
$m := r-l$ \tcp*{\# of low elements between $L$ and $R$}\;
  \lIf{ $\rank \leq  m $ }{  \Return Query$(A$.low, $L$, $R$, $\rank)$}\;  %
  \lElse{\hspace*{1.75cm}  \Return Query$(A$.high, $L$, $R$, $\rank - m)$} %

\end{algorithm}

Algorithm~\ref{alg:query} gives pseudo-code for the query,
which performs preprocessing (i.e., splitting the array into two smaller arrays) 
only where needed. Note that we have to keep three things separate here: values that
are relevant for median computation and partitioning the input, positions in the input sequence
that are relevant for finding the elements within the range $[L,R]$, and positions in 
the subdivided arrays that are important for counting elements.

Let us first analyze the time required for processing a query not
counting the `preprocessing' time within lines 4--6: The query
descends $\log_2 n$ levels of recursion. On each level,
\Id{Find}-operations for $L$ and $R$ are performed on the lower half
of the current subproblem.  If we  used binary search, we would
get a total execution time of up to $\sum_{i=1}^{\log_2 n}
O(\log\frac{n}{2^i})=\Th{\log^2 n}$.  However, the fact that in all
these searches, we search for the same key ($L$ or $R$) allows us to use
a standard technique called Fractional Cascading
\cite{fractionalcascading} that reduces the search time to a constant,
once the position in the first search is known. Indeed, we only need a rather basic
variant of Fractional Cascading, which applies when each
successor list is a sublist of the previous one \cite{BKOS97}.  Here,
it suffices to augment an element $e$ of a list with a pointer to the
position of some element $e'$ in each subsequent list (we
have two successors -- $A.\Id{low}$ and $A.\Id{high}$).  In our case, we need
to point to the largest element in the successor that is no larger
than $e$.  We get a total search time of $O(\log n)$.

Now we turn to the preprocessing code in lines~4--6 of
Algorithm~\ref{alg:query}.  Let $s(i)$ denote the level of recursion
at which query~$i$ encountered an undefined array $A.low$ for the
first time. Then the preprocessing time performed during query $i$ is
$O(n/2^{s(i)})$ if a linear time algorithm is used for median
selection \cite{DBLP:conf/stoc/BlumFPRT72} (note that we have a linear
recursion with geometrically decreasing execution times).
This preprocessing time also includes the cost of 
finding the pointers for Fractional Cascading while splitting the list in lines~4--6.
Since the preprocessing time during query $i$
decreases with $s(i)$, the total preprocessing time is maximized if
small levels $s(i)$ appear as often as possible.  However, level~$j$
can appear no more than $2^j$ times in the
sequence $s(1),s(2),\ldots, s(k)$.%
\footnote{Indeed, for $j>0$ the maximal number is $2^{j-1}$ since the other half of the available subintervals have already been covered by the preprocessing happening in the layer above.}  Hence,
we get an upper bound for the preprocessing time when the smallest $\floor{\log k}$
levels are used as often as possible (`filled') and the remaining
levels are $\ceil{\log k}$.  The preprocessing time at every used
level is $O(n)$ giving a total time of $O(n\log k)$. The same bound
applies to the space consumption since we never allocate memory that
is not used later.
We summarize the main result of this section in a theorem:
\begin{theorem}
  The online range median problem (RMP) on an array with $n$ elements
  and $k$ range queries can be solved in time $O(n\log k + k\log n)$  and
  space $O(n\log k)$.
\end{theorem}

Another variant of the above algorithm invests $O(n\log n)$ time and space
into complete preprocessing up front. Subsequently,  any range median query can be answered in $O(\log n)$ time.
This improves the preprocessing
space of the corresponding result in \cite{DBLP:journals/njc/KrizancMS05} by a factor $\log n/\log\log n$ and the preprocessing time by at least this factor.

\subsection{Lower Bounds}
\label{ss:lowerbound}

We shortly discuss how far our algorithm is from optimality.
In \cite{DBLP:conf/esa/Har-PeledM08}, 
a comparison-based lower bound of $\Omega(n \log k)$ is shown for the range median problem 
\footnote{The authors derive a lower bound of $l :=\frac{n!}{k!((n/k-1)!)^k}$, where $n$ is a multiple of $k<n$. Unfortunately, the analysis of the asymptotics of $l$ given in \cite{DBLP:conf/esa/Har-PeledM08} is erroneous; however, a corrected analysis shows that the claimed $\Omega(n\log k)$ bound holds.}. 
As our algorithm shows, this bound is (asymptotically) tight if $k \in O(n)$.
For larger $k$, the above lower bound is no longer valid, as the construction requires $k<n$. 
Yet, a lower bound of $\Omega(n\log n)$ is immediate for $k\geq n$, considering only the first $n-1$ queries.
Furthermore, $\Omega(k)$ is a trivial lower bound.
Note that in the analysis of our algorithm, the number of levels of the recursion is
actually bounded by $O(\min \{\log k, \log n \})$, and 
thus
for any $k\geq n$ our algorithm has running time $O(n\log n + k\log n)$,
which is up to $\Omega(\log n)$ from optimal for any $k$.

In a very restricted model (sometimes called ``Pointer Machine''), where a memory location can be reached only by following pointers, and not by direct addressing, 
our algorithm is indeed optimal also for $k \geq n$: it takes $\Omega(\log n)$ time to even access an arbitrary element of the input.
Since every element of the input is the answer to at least one range query (e.g. the query whose range contains only this element), the bound follows.
An interesting question is whether a lower bound $\Omega(k\log n)$ could be shown in stronger models. However, note that any comparison-based lower bound (as the one in  \cite{DBLP:conf/esa/Har-PeledM08}) cannot be higher than $\Omega(n\log n)$: With $O(n\log n)$ comparisons, an algorithm can determine the permutation of the array elements, which suffices to answer any query without further element comparisons.
Therefore, one would need to consider stronger models (e.g. the ``cell-probe'' model), in which proving
lower bounds is significantly more difficult.

\section{A Linear Space RAM Implementation}\label{s:RAM}

\begin{algorithm}%
\caption{Query($A,L,R,\rank$)} \label{alg:linearquery}
\dontprintsemicolon
{\bf Input:} range select data structure $A$, query range $[L,R]$ and desired rank $\rank$ \;
\lIf{$|A|=1$}{\Return $A[1]$} \;
\uIf{$A.{low}$ is undefined}{
  Compute median $x$ value of the values in $A$ \;
  $A$.lowbits := BitVector$(|A|,\setGilt{i\in 1..|A|}{A[i]\leq x})$ \;
  $A$.low\hspace*{1.5mm}    := $\seqGilt{A[i]}{i\in 1..|A|, A[i]\leq x}$ \;
  $A$.high    := $\seqGilt{A[i]}{i\in 1..|A|, A[i]> x}$ \;
  deallocate the value array of $A$ itself
}
$ l := $ $A$.lowbits.rank$(L-1)$  \;
$ r    := $ $A$.lowbits.rank$(R)$  \;
$ m    := r-l$\\
\lIf{ $\rank \leq  m $ }{  \Return Query$(A$.low, $l+1$, $r$, $\rank)$}\;  %
\lElse{\hspace*{1.75cm}  \Return Query$(A$.high, $L-l$, $R-r$, $\rank - m)$} %
\end{algorithm}

Our starting point for a more space efficient implementation of
Algorithm~\ref{alg:query} is the observation that we do not actually
need all the information available in the arrays stored at the
interior nodes of our data structure. All we need is support for the
operation \Id{Find}$(x)$ that counts the number of elements $e$ in
$A.low$ that have index $e.i\leq x$.  This information can already be
obtained from a bit-vector where a $1$-bit indicates whether an
element of the original array  is in $A.low$. For this bit-vector, the operation
corresponding to \Id{Find} is called \Id{rank}.  In the RAM model, there are data
structures that need space $n+o(n)$ bits,  can be constructed in
linear time and support \Id{rank} in constant time (e.g.,
\cite{Cla88,OkaSad06}\footnote{Indeed, since we only need the rank operation, there are very simple and efficient implementations: store a table with ranks for
indices that are a multiple of $w=\Theta(\log n)$. General ranks are then the sum of the next smaller table entry and the number of $1$-bits in the bit array between this rounded position and
the query position. Some processors have a POPCNT instruction for this purpose. Otherwise we can use lookup tables.}).  Unfortunately, this idea alone is not enough
since we would need to store $2^j$ bit arrays consisting of $n$
position each on level $j$.  Summed over all levels, this would still
need $\Omega(n\log^2 n)$ bits of space even if optimally compressed data
structures were used.  This problem is solved using an additional
idea: for a node  of our data structure  with value array $A$, we do not store a bit
array with $n$ possible positions but only with $|A|$ possible
positions, i.e., bits represent positions in $A$ rather than in the
original input array. This way, we have $n$ positions on every
\emph{level} leading to a total space consumption of $O(n\log n)$ 
bits.  For this idea to work, we need to be able to transform the
query range in the recursive call in such a way that \Id{rank}
operations in the contracted bit arrays are meaningful. Fortunately,
this is easy because the rank information we compute also defines the
query range in the contracted arrays. Algorithm~\ref{alg:linearquery}
gives pseudocode specifying the details.  Note that the algorithm is
largely analogous to Algorithm~\ref{alg:query}.  In some sense, the
algorithm becomes simpler because the distinction between query
positions and array positions for counting disappears (If we still
want to report the positions of the median values in the input, we can
store this information at the leaves of the data structure using
linear space). Using an analysis analogous to the analysis of
Algorithm~\ref{alg:query}, we obtain the following theorem:

\begin{theorem}
  The online range median problem (RMP) on an array with $n$ elements
  and $k$ range queries can be solved in time $O(n\log k + k\log n)$ and
  space $O(n)$ words on the RAM model.
\end{theorem}

By doing all the proprocessing up front, we obtain an algorithm
with preprocessing time $O(n\log n)$ using $O(n)$ space and query time $O(\log n)$.
This improves the space consumption compared to \cite{DBLP:journals/njc/KrizancMS05} by a factor $\log^2 n/\log\log n$. 

\section{Higher Dimensions}\label{s:highd}

Since our algorithm decomposes the values rather than the positions of
elements, it can be naturally generalized to higher dimensional point
sets. We obtain an algorithm that needs $O(n\log k)$ preprocessing
time plus the time for supporting range counting queries on each
level.  The amortized query time is the time for $O(\log n)$ range
counting queries.  Note that query ranges can be specified in any way
we wish: (hyper)-rectangles, circles, etc., without affecting the way
we handle values.  For example, using the
data structure for 2D range counting from \cite{DBLP:conf/isaac/JaJaMS04} we obtain a
data structure for the 2D rectangular range median problem that needs
$O(n\log n\log k)$ preprocessing time, $O(n\log k/\log\log n)$ space, and
$O(\log^2n/\log\log n)$ query time. This not only applies to 2D arrays consisting
of $n$ input points put to arbitrary two-dimensional point sets with
$n$ elements.  

Unfortunately, further improvements, e.g. to logarithmic query time
seem difficult. Although the query range is the same at all levels of
recursion, Fractional Cascading becomes less effective when the result of a
rectangular range counting query is defined by more than a constant
number of positions within the data structure because we would
have to follow many forwarding pointers.  Also, the
array contraction trick that allowed us to use 
dense bit arrays
in Section~\ref{s:RAM} does not work anymore because an array with
half the number of bits need not contain any empty rows or columns.

Another indication that logarithmic query time in two dimensions might
be difficult to achieve is that there has been intensive work on the
more specialized median-filtering problem in image processing where we
ask for \emph{all} range medians with query ranges that are squares of
size $2r+1\times 2r+1$ in an image with $n$ pixels. The best previous algorithms known here need
time $\Th{n\log^2 r}$\cite{GilWerman} unless the range of values is very small
\cite{2Dmedian1,2Dmedian2}. Our result above improves this by a factor $\log\log r$ 
(by applying the general algorithm to input pieces of size $3r\times 3r$) but this
seems to be of theoretical interest only.

\section{Dynamic Range Medians}
\label{s:dynamic}

In this section, we consider a dynamic variant of the RMP,
where we have a linked list instead of an array,
and elements can be deleted or inserted arbitrarily.
In this setting, we still want to answer median queries, whose range is given by two pointers to the first and the last
element in the query range. 

In the following, we sketch a solution which allows  inserts and deletes   in $O(\log^2 n)$ amortized time each, and range median queries in $O(\log^2 n)$ worst case time. 
The basic idea is to use a BB($\alpha$) tree \cite{DBLP:conf/stoc/NievergeltR72} as a primary structure, in which all elements are ordered by their \emph{value}.
With each inner node, we associate a secondary structure, which contains all the elements of the node's subtree, ordered by their position in the input list.
More precisely, we store these elements in a balanced binary search tree, where nodes are augmented with a field indicating the size of their subtree, see e.g. \cite{DBLP:conf/icalp/Roura01}. 
This data structure permits to answer a range query by a simple adaptation of Algorithm~\ref{alg:query}: starting at the root,
we determine the number of elements within the query range which are in the left subtree, and depending on the result continue the search for the median in the left or in the right subtree. The required counting in each secondary structure takes $O(\log n)$ time, and we need to perform at most $O(\log n)$ such searches for any query.
When an element is inserted or deleted, we follow the search path in the BB($\alpha$) tree according to its value, and update all the $O(\log n)$ secondary structures of the visited nodes. 
The main difficulty arises when a rotation in the BB($\alpha$) tree is required: in this case, the secondary structures are rebuilt from scratch, which costs $O(p\log p)$ time if the subtree which is rotated contains $p$ nodes. 
However, as shown in \cite{mehlhornsbook,luekerwillard}, such rotations are required so rarely that 
the amortized time of such an event is only $O(\log p \log n)=O(\log^2 n)$.

We note that using this dynamic data structure for the one-dimensional RMP, we can implement a two-dimensional median filter, by scanning over the image, maintaining all the pixels in a strip of width $r$. In this way, we obtain a running time of $O(\log^2 r)$ per pixel, which matches the state-of-the-art solution for this problem \cite{GilWerman}.
This indicates that obtaining a solution with $O(\log n)$ time for all operations could be difficult.

\section{Conclusion}\label{s:conclusions}

We have presented improved upper bounds for the range median problem. 

The query time of  our solution is asymptotically optimal for $k \in O(n)$, or when all preprocessing has to be done up front.
For larger values of $k$, our solution is at most a factor $\log n$ from optimal. In a very restricted model where no arrays are allowed, our solution is optimal for all $k$. Moreover, in the RAM model, our data structure requires only $O(n)$ space, which is clearly optimal. Making the data structure dynamic adds a factor $\log n$  to the query time.
Using $O(n^2)$ space, it is trivial to precompute all medians of a given array so that the query time
becomes constant. However, it is open whether the term $O(k \log n)$ in the query time could be reduced towards $O(k)$ in the RAM model when $k=o(n^2)$.

Given the simplicity of our data structure, a practical implementation would be easily possible.
To avoid the large constants involved when computing medians for recursively splitting the array,
one could use a pivot chosen uniformly at random. This should work  well in expectation.

It would be interesting to find faster solutions for the dynamic RMP or the two-dimensional (static) RMP:
Either would lead to a faster median filter for images, which is a basic tool in image processing.

\bibliographystyle{alpha}
\bibliography{references}

\end{document}